\documentclass[sigplan,font=10pt]{acmart}
\acmConference[VMIL'19]{ACM SIGPLAN International Workshop on Virtual Machines and Intermediate Languages}{October 22, 2019}{Athens, Greece}
\acmYear{2019}
\acmISBN{} % \acmISBN{978-x-xxxx-xxxx-x/YY/MM}
\acmDOI{} % \acmDOI{10.1145/nnnnnnn.nnnnnnn}
\startPage{1}

\setcopyright{none}
\usepackage{tikz}
\usepackage{listings}
\usepackage{subcaption}
\hypersetup{draft}

\newcommand{\ie}{\textit{i.e.,}}

\lstdefinestyle{grace}
{
  language=java,
  basicstyle=\fontsize{9}{9}\selectfont\tt,
  keywordstyle=\fontsize{10}{10}\selectfont\tt\bfseries,
  commentstyle=\fontfamily{ptm}\selectfont\itshape\color{magenta},
  numbers=none,
  numberstyle=\tt\footnotesize,
  literate={~~>}{$\leadsto\ $}{2} {lambda}{$\lambda$}{0} {<<}{$\llbracket$}{1}
               {>>}{$\rrbracket$}{1} {->}{$\to$\ }{1},
  aboveskip=1ex,
  belowskip=1ex,
  tabsize=1,
  columns=fullflexible,
  xleftmargin=0ex,
  resetmargins=false,
  showstringspaces=false,
  escapeinside={(*@}{@*)},
  morecomment=[l]{--},
  backgroundcolor=\color{white},%
morekeywords={trait,async,spawn,get,this,string,uint,int,real,bool,var,val,then,fun,match,
                          lin,imm,local,borrowed,Unit,method,linear,rec,send,moved,lent,?,local,self,class}
}

\newcommand{\ec}[1]{\lstinline[style=grace,basicstyle=\tt]@#1@}
\newcommand{\kw}[1]{\text{\ec{#1}}}

\newcommand{\async}[1]{\ensuremath{\kw{spawn}~ #1}}
\newcommand{\rec}[0]{\ensuremath{\kw{rec}}}

\newcommand{\send}[2]{\ensuremath{\kw{send} ~#1 \leftarrow #2}}
\newcommand{\new}[2]{\ensuremath{#1(#2)}}
\newcommand{\err}[0]{\ensuremath{\kw{Error}}}
\newcommand{\loc}[0]{\ensuremath{l}}

\newcommand{\linkw}[0]{\ensuremath{\kw{lin}}}

\newcommand{\localkw}[0]{\ensuremath{\kw{local}}}

\newcommand{\lin}[1]{\ensuremath{\linkw~ #1}}

\newcommand{\local}[1]{\ensuremath{\localkw~ #1}}
\newcommand{\bor}[1]{\ensuremath{\bor}~ #1}

\newcommand{\moved}[0]{\ensuremath{\kw{moved}}}
\newcommand{\lent}[0]{\ensuremath{\kw{lent}}}

\newcommand{\movable}[0]{\ensuremath{\kw{movable}}}
\newcommand{\unmov}[0]{\ensuremath{\kw{immov}}}

\newcommand{\unit}[0]{\ensuremath{\kw{Unit}}}

\newcommand\lrog{\mathrm{mrog}_{\sigma}(l)}
\newcommand{\rog}[1]{\mathrm{mrog}_{\sigma}(#1)}
\newcommand{\updateStore}[3]{{#2}_{#3}(#1)}

\newcommand{\ntyperule}[3]{
\begingroup
    \fontsize{8pt}{12pt}\selectfont
  \begin{array}{c}
    \textsc{\scriptsize ({#1})} \\
    #2 \\
    \hline
    \raisebox{-1pt}{$#3$}
  \end{array}
\endgroup
}

\newcommand{\nreduction}[2]{
\begingroup
    \fontsize{8pt}{12pt}\selectfont
  \begin{array}{c}
    \textsc{\scriptsize (#1)} \\
    #2
  \end{array}
\endgroup
}

\begin{document}

\title{Towards Gradual Checking of Reference Capabilities}
                                        %% when present, will be used in
                                        %% header instead of Full Title.
%\titlenote{with title note}             %% \titlenote is optional;
                                        %% can be repeated if necessary;
                                        %% contents suppressed with 'anonymous'
\subtitle{Work in Progress}                     %% \subtitle is optional
%\subtitlenote{with subtitle note}       %% \subtitlenote is optional;
                                        %% can be repeated if necessary;
                                        %% contents suppressed with 'anonymous'

%% Author information
%% Contents and number of authors suppressed with 'anonymous'.
%% Each author should be introduced by \author, followed by
%% \authornote (optional), \orcid (optional), \affiliation, and
%% \email.
%% An author may have multiple affiliations and/or emails; repeat the
%% appropriate command.
%% Many elements are not rendered, but should be provided for metadata
%% extraction tools.

\author[K. Fernandez-Reyes]{Kiko Fernandez-Reyes}
\affiliation{
  \institution{Uppsala University}
  \country{Sweden}
}
\email{kiko.fernandez@it.uu.se}

\author[Isaac O. G.]{Isaac Oscar Gariano}
\affiliation{
  \institution{Victoria University of Wellington}
  \country{New Zealand}
}
\email{isaac@ecs.vuw.ac.nz}

\author[J. Noble]{James Noble}
\affiliation{
  \institution{Victoria University of Wellington}
  \country{New Zealand}
}
\email{kjx@ecs.vuw.ac.nz}

\author[T. Wrigstad]{Tobias Wrigstad}
\affiliation{
  \institution{Uppsala University}
  \country{Sweden}
}
\email{tobias.wrigstad@it.uu.se}

%\authorrunning{K. Fernandez-Reyes, I. Gariano, J. Noble, T. Wrigstad}
%% Abstract
%% Note: \begin{abstract}...\end{abstract} environment must come
%% before \maketitle command
\begin{abstract}
Concurrent and parallel programming is difficult due to the presence
of memory side-effects, which may introduce data races. Type qualifiers, such as
reference capabilities, can remove data races by restricting sharing of mutable data.

Unfortunately, reference capability languages are an all-in or nothing game,
\ie{} all the types must be annotated with reference capabilities.
In this work in progress, we propose to mix the
ideas from the reference capability literature with gradual typing,
leading to gradual reference capabilities.
\end{abstract}

%% 2012 ACM Computing Classification System (CSS) concepts
%% Generate at 'http://dl.acm.org/ccs/ccs.cfm'.
%\begin{CCSXML}
%<ccs2012>
%<concept>
%<concept_id>10011007.10011006.10011008</concept_id>
%<concept_desc>Software and its engineering~General programming languages</concept_desc>
%<concept_significance>500</concept_significance>
%</concept>
%<concept>
%<concept_id>10003456.10003457.10003521.10003525</concept_id>
%<concept_desc>Social and professional topics~History of programming languages</concept_desc>
%<concept_significance>300</concept_significance>
%</concept>
%</ccs2012>
%\end{CCSXML}

%\ccsdesc[500]{Software and its engineering~General programming languages}
%\ccsdesc[300]{Social and professional topics~History of programming languages}
%% End of generated code

%% Keywords
%% comma separated list
  %% \keywords are mandatory in final camera-ready submission

%% \maketitle
%% Note: \maketitle command must come after title commands, author
%% commands, abstract environment, Computing Classification System
%% environment and commands, and keywords command.
\maketitle

\section{Introduction}
Data-races are one of the core problems that makes concurrent and
parallel programming difficult. Let us illustrate the problem
with an implementation
of a \verb|collection| class (borrowed from Grace's library~\cite{grace}).

\begin{lstlisting}[style=grace,xleftmargin=2ex]
class collection<<T>> {
  var iterator
  method first {
    def it = self.iterator
      if (it.hasNext) then {
        it.next        
      } else {
        stdGrace.BoundsError.raise "no first element" 
      }}}
\end{lstlisting}

This class is not thread-safe: when two threads have access
to the same instance and one thread reads and the other writes that instance,
data races may occur, specially in the presence of
synchronisation on variables~\cite{Peterson,PetersonFailed}. Using manual
synchronisation mechanisms such as locks and monitors
can fix this problem, but can exacerbate it by causing deadlocks.

Static reference capability programming languages
eliminate data-races at compile-time, by adding extra type annotations which
place constraints on references~\cite{CapabilityBook,SFMEncore,EncoreTS,PonyTS,Gordon,RelaxedLinear}.
Taking inspiration from the capability-based Encore language \cite{SFMEncore,EncoreTS}
and reusing the example above, one could add reference capabilities to the
\verb|collection| class to either forbid sharing or making the collection alias-free
as follows,
\begin{lstlisting}[style=grace,numbers=left,xleftmargin=8ex]
local class collection<<local T>> (*@\label{code:localDecl}@*)
{  var iterator : local (*@\label{code:localVar}@*) }

linear class collection<<t>>(*@\label{code:linearClass}@*)
\end{lstlisting}

In this example, the developer forbids sharing the class by using the \local{} type qualifier at the class,
type parameter, and field declaration (lines~\ref{code:localDecl}--\ref{code:localVar}).
Class declarations annotated with a \ec{linear} type qualifier (line \ref{code:linearClass})
forbid aliasing, so that any attempt at creating an alias is rejected by the type checker.

Reference capability-based type systems guarantee data-race freedom
at the expense of annotating the whole program.
In this work, we investigate gradually adding reference capabilities
to an untyped language, leading to a gradual reference capability-based language.
We argue that gradual
reference capabilities are orthogonal to the typing discipline, making our approach
suitable for both statically-typed and gradually-typed languages.
Our end goal is to introduce gradual reference capabilities to a gradually typed language.

%\begin{figure}[t]
%  \includegraphics[scale=0.38,trim=3cm 15cm 10cm 3cm,clip=true]{image/image.png}
%  \caption{\label{fig:end goal}Towards a gradually capable typed language}
%\end{figure}

%Info: 4 pages excluding biblio. Deadline August 29, 2019
%
%Gradual typing as a natural step.
%Static type systems enforce object properties.
%Languages and their approach to adding linear types.
%Reference capabilities at the other extreme.
%Finish with explanation on gradual types and why not gradual capabilities?

%\paragraph{Contributions}

\section{Overview}

In an untyped, actor-based language with shared memory, actors may send
objects between themselves in an unsafe way. For example, the following
code shows an actor sending a message to another actor,
which contains the reference of a file handle, and closing the file immedialy after sending~it:

\begin{lstlisting}[style=grace,xleftmargin=8ex]
fileHandle = open("...")
otherActor.send(fileHandle)
fileHandle.close()
\end{lstlisting}

When the second actor receives the handle, it could still write to the file
if the actor that closes the file is not fast enough, leading to a data-race,
or they could both close the file, leading to a runtime error.

\begin{figure}[t!]
%    \centering
%    \begin{subfigure}[t]{0.4\textwidth}
%        \centering 
        \includegraphics[height=1in,trim=0cm 15cm 23cm 1cm,clip=true]{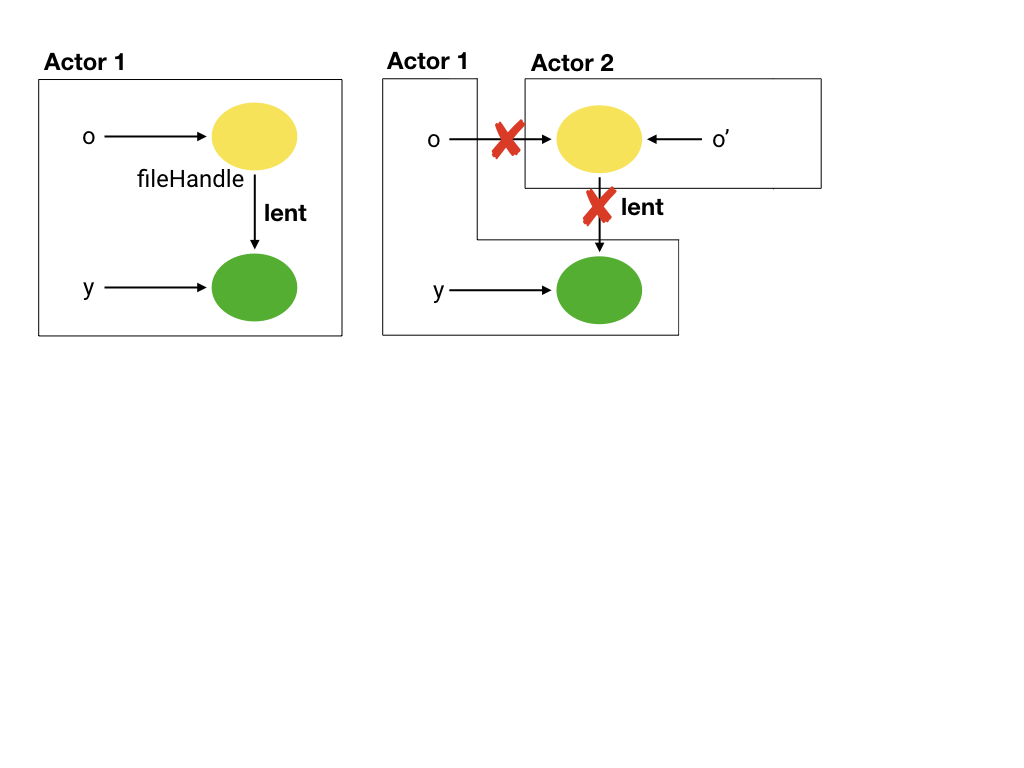}
        \caption{\label{fig:actor1}\ec{Actor1} has two references to objects, where the yellow object has a reference to the green object.}
%    \end{subfigure}%
%    \qquad\qquad
%    \begin{subfigure}[t]{0.4\textwidth}
%        \qquad\qquad\qquad\!
%        \includegraphics[height=1in,trim=13cm 15cm 0cm 1cm,clip=true]{image/image2.png}
%        \caption{\label{fig:actor2}\ec{Actor1} has sent the reference \ec{x} to \ec{Actor2}. Sending an object that borrows
%        references uninitialises the borrowed references.}
%    \end{subfigure}   
%    \caption{\label{fig: picto}Pictographical representation of an actor that sends an object that borrows a reference to another actor.}
\end{figure}

To address the need for sharing mutable data in an untyped language we introduce
gradual referential capabilities to the language. Developers can
add reference capabilities to enforce their desired behaviour, but without
having to annotate the whole program. For instance,
they could use the \moved{} capability for safely sharing an object between actors:

\begin{lstlisting}[style=grace,xleftmargin=8ex]
moved fileHandle = open("...")
otherActor.send(fileHandle)
fileHandle.close()
\end{lstlisting}

\noindent
A \moved{} capability ensures that the \ec{fileHandle} object along with the transitive closure of
all of its (movable) reachable references, change their ownership from the current actor to the \ec{otherActor}.
The example above could be rejected either statically or dynamically
since the \ec{fileHandle} cannot be used again after it has been sent.
The semantics are the same as \ec{Transferable} objects in JavaScript~\cite{JavaScriptTransferable}, except that
JavaScript works with \ec{Workers} instead of actors.
Alternatively, developers could use a \lent{} capability to \textit{borrow} references
but forbid sending the reference to another actor:

\begin{lstlisting}[style=grace,xleftmargin=8ex]
lent fileHandle = open("...")
otherActor.send(fileHandle)
fileHandle.close()
\end{lstlisting}

\noindent
The current actor borrows the \ec{fileHandle} reference,
and the type checker and runtime forbid sharing the \ec{fileHandle} reference
to another actor. The example above
is rejected statically during type checking, or by the runtime which
prevents a \lent{} annotated reference from being sent to
another actor. A more complex case considers that an actor has
an object that borrows a reference in a field~(Figure~\ref{fig:actor1}):

\begin{lstlisting}[style=grace,xleftmargin=8ex]
lent fileHandle = open("...")
var o = new Object(fileHandle)
otherActor.send(o)
fileHandle.close()
\end{lstlisting}

\noindent
Sending object \ec{o} to \ec{otherActor} is not allowed, since the
transitive closure of its references must be movable, which results in
a runtime error.

These two examples exercise different aspects of the type system
and runtime, \ie{} the permission to share an object and the restriction
to not escape the current actor once a reference is borrowed.

\section{Formal Semantics}
\begin{figure}[t]
 \begin{align*}
 \textit{Classes} & &
   \textit{CL} & ::= \textbf{class}\  C(\overline{f : \kappa\ })\{ \overline{M} \} \\
  \textit{Methods} & &
   M &::= \kappa\ \textbf{method}\ m(\overline{x : \kappa\ }) \to \kappa\  \{\ e\ \} \\
 \textit{Expressions} & &
   e &::= e.m(\overline{e}) \mid e.f \mid e.f := e \\
   &&& \mid \kappa\ x := e; e \mid \new{C}{\overline{e}} \mid \async{e}\\
   & & & \mid \rec \mid \send{e}{e} \mid v \mid x\\
  \textit{Values} & &
   v &::= \unit \mid a \mid\ \rho\ \loc \mid \rho\ \err \\
 \textit{Capabilities}  & &
   \kappa & ::=\ ? \mid \moved{} \mid \lent{} \\
\textit{Permission}  & &
   % @ to \movable or similar same for \unmov
   \rho & ::= \movable \mid \unmov\\ % runtime notation for not-borrowed and borrowed.
 \textit{Evaluation Context} & &
   E & ::= \bullet \mid E.m(\overline{e}) \mid v.m(\overline{v}, E, \overline{e}) \mid E.f \\
   & & & \mid e.f := E \mid x := E; e \mid \new{C}{\overline{v},E,\overline{e}} \\
   & & & \mid \send{E}{e} \mid \send{v}{E} \\
 \textit{Store} & &
  & \sigma : \begin{cases}
  &x  \to v \\
  & l  \to \new{C}{\overline{v_i}}
                   \end{cases} \\
 \textit{Class Table} & &
  & \Delta : C \to CL
 \end{align*}
\caption{\label{fig: language}
Syntax of language. $C$, $m$, $f$, $x$, and $t$ are meta-variables representing class, method, fields, variable names
and actor ids; $?$ represents the dynamic capability.}
\end{figure}

We define a concurrent, untyped object-oriented calculus
with gradual reference capabilities~(Fig.~\ref{fig: language}). Meta variables $C$, $m$, $f$, $x$, and $a$ range over
class, method, field, variable names, and actor ids.
A class has a name $C$,
followed by field declarations and method declarations. Field declarations
($\overline{f : \kappa}$) have capability $\kappa$;
method declarations have capability $\kappa$ applied to the
implicit reference \ec{this},
name $m$, parameters $\overline{x}$ with capability $\kappa$
and returns a capability $\kappa$ applied to resulting value.
Expressions are method calls ($e.m(\overline{e})$), field accesses
($e.f$), assignment to field and variables ($e.f := e; e$ and $ \kappa\ x := e; e$), creation of new instances ($\new{C}{\overline{e}}$), spawning of a new actor ($\async{e}$),
and receiving and sending messages ($\rec$ and \send{e}{e}).
Values are constants, variables, unit and actor ids;
runtime locations ($\rho \ \loc$) and errors ($\rho\ \err$) are not part of the surface syntax. Available capabilities are the
dynamic capability $?$ (which represents the omission of
a capability), and the \moved{} and \lent{} capabilities for allowing
thread-sharing of data and borrowing references that forbid
sharing an object outside the current actor, respectively. The runtime keeps track of
these capabilities by tracking their usage ($\rho$),
where $\movable$ represents that the reference can be passed to another actor
and $\unmov$ that the reference cannot be passed to another actor.

The operational semantics are based on small-step, reduction-context based rules for evaluation within actors. The 
evaluation context contains a hole $\bullet$ that denotes where the next reduction happens~\cite{EvContext}.
The runtime semantics~(Fig.~\ref{fig: runtime}) have an input store $\sigma$ and an output store $\sigma'$, the store also contains the state of all the actors (their message queues and main expression); these actors will execute concurrently, the runtime will arbitrarily choose a non-blocked actor to execute next.
We denote such an actor $a$ with pending messages $\overline{v}$ currently executing $e$ as

\[a \mapsto \overline{v}\ E[e]\]

For clarity, we omit concurrency details 
whenever they are not relevant for the current evaluation, such that:

\begin{align*}
(\sigma, a \mapsto \overline{v}\ E[e]) &\to (\sigma', a \mapsto \overline{v}\ E[e']) \\
\text{ where } \sigma | e &\to \sigma' | e', \textit{ and } \err \notin e
\end{align*}

Definition~\ref{def: store} states the store modifications
when a value is used as a capability $\kappa$. Casting an \unmov{} 
reference to be \moved{} throws a runtime error, \textit{i.e.,} 
$\updateStore{\sigma}{\moved{}}{\unmov\ \loc} = \textit{undefined}$. The most interesting
case is when a \movable{} location $\loc$ is cast to \moved{}:
for a location \loc{} to be moved the transitive closure of the reachable object graph
must be movable ($\textit{rog}_\sigma(\loc{}) = \lrog$ and Definition~\ref{def: rog}).
For all other locations,
uninitialise all references that reach to the objects that are going to be moved.
Variables whose locations point to the movable reachable object graph of the
object sent are also uninitialised (\textit{i.e.,} case $\sigma'(x)$).

Definitions~\ref{def: rrog}--\ref{def: rog}
define the reachable object graph (\textit{ROG}) and the movable \textit{ROG},
which traverses through locations and \movable\ locations, respectively. 

Definition~\ref{def: cast} casts a value to a given capability $\kappa$, returning an annotated value that the runtime 
keeps track of. For example, trying to cast a \unmov{}
reference to a \movable{} reference throws an error. These casts happen at runtime and are not available in the surface language.

%\begin{figure}
\begin{definition}\label{def: store}
Define $\updateStore{\sigma}{\kappa}{v}$ to modify the
store $\sigma$ whenever $v$ is used as capability $\kappa$:
\begin{align*}
\updateStore{\sigma}{\kappa}{v}&= \sigma, \text{ if } \kappa \neq\moved{} \\
\updateStore{\sigma}{\moved{}}{\unmov\ \loc} & = \textit{undefined}\\
\updateStore{\sigma}{\moved}{\movable\ \loc} & = \sigma' \text{ where }\\
\textit{rog}_\sigma(\loc{}) &= \lrog \text{ and }
 \forall \loc', x, a:\\
m &= \lrog \\
\sigma'(\loc') &=
\begin{cases}
  \sigma(\loc'),&
   \text{if } l' \in m \\ % complement operation
   &\\
  \sigma(l')[\overline{m := \err}],& \text{otherwise}
\end{cases}\\
\sigma'(x) &= \begin{cases}
  \rho~\err,& \text{ if } \sigma(x) = \rho\ \loc' \text{ and }  \loc' \in\ m\\
  \sigma(x),& \text{otherwise}
\end{cases}\\
\sigma'(a) &= \sigma(a)
\end{align*}
\end{definition}

\begin{definition}\label{def: rrog}
Define $\textit{rog}_\sigma(v)$ to be the reachable object graph of $v$:
\begin{align*}
\textit{rog}_\sigma(\rho\ \loc) &= \{ \loc \} \overline{\bigcup \textit{rog}_\sigma(v)}, \text{ where } \sigma(\loc) = C\{\overline{v}\} \\
\textit{rog}_\sigma(v) &= \varnothing, \text{ otherwise}
\end{align*}
\end{definition}

\begin{definition}\label{def: rog}
Define $\rog{v}$ to be the movable reachable object graph of $v$:
\begin{align*}
\rog{\movable\ \loc} &= \{ \loc \} \overline{\bigcup \rog{v}}, \text{ where } \sigma(\loc) = C\{\overline{v}\} \\
\rog{v} &= \varnothing, \text{ otherwise}
\end{align*}
\end{definition}

\begin{definition}\label{def: cast}
Define $\kappa(v) = v'$, to cast $v$ to $\kappa$:
\begin{align*}
\lent(\movable\ \loc) &= \unmov\ \loc \\
\moved(\unmov\ \loc) &= \textit{undefined} \\
\kappa(v) &= v
\end{align*}
\end{definition}

%\caption{\label{fig: defs}Definitions of movable reachable object graph, casting, and store modification.}
%\end{figure}
% TODO:
% - Let-reduction rule -- done
% - Type setting moved in Def 3.3 -- done
% - Change t to a -- DONE
% - Change @ and & such as moveable and immovable -- DONE
% - Explain definitions -- DONE
% - Add example -- DONE
% - We have lent instead of local -- DONE
% - method explanation -- DONE
% - Overall pass

\newcommand{\NewClass}[0]{E-NewClass}
\newcommand{\FieldA}[0]{E-FieldAccess}
\newcommand{\Err}[0]{E-Error}
\newcommand{\Assign}[0]{E-Assignment}
\newcommand{\Async}[0]{E-Spawn}
\newcommand{\MethodC}[0]{E-MethodCall}
\newcommand{\Recv}[0]{E-Receive}
\newcommand{\Send}[0]{E-Send}
\newcommand{\Variable}[0]{E-Variable}
\newcommand{\Let}[0]{E-VarAssignment}

\begin{figure*}[t]
\centering
$$
%\nreduction{E-Abbrev}{
%\sigma, t \mapsto \overline{v}\ E[e] \to \sigma', t \mapsto \overline{v}\ E[e'] \\
%\text{ where } \sigma | e \to \sigma' | e'
%}
%\qquad
\ntyperule{\NewClass}
{\Delta(C) = \textbf{class} C(\overline{\kappa f}) \{ \overline{M} \}}
{\sigma | C(\overline{v}) \to \overline{\updateStore{\sigma}{\kappa}{v}}, l \mapsto \new{C}{\overline{\kappa(v)}} | \movable{}\ \loc{}}
\qquad
\ntyperule{\Let}
{x' \text{ is fresh}}
{
\sigma | \kappa\ x := v; e \to \updateStore{\sigma}{\kappa}{v}, x' \mapsto \kappa(v) | e[x := x']
}
\qquad
\ntyperule{\FieldA}
{
\sigma(l) = \new{C}{\overline{f = v}}
}
{
\sigma | \rho\ l.f_i \to \sigma | v_i
}
$$
$$
\ntyperule{\Assign}
{  \sigma(\loc) = \new{C}{\overline{f = v'}},\quad
  \updateStore{\sigma}{\kappa_i}{v} = \sigma', \\
  \Delta(C) = \textbf{class}\ C(\overline{\kappa\ \loc}) \{ \ldots \}\\
  \sigma'(\loc) = \new{C}{\overline{v''}}, \quad 
  \forall j \neq i, v'''_j = v''_j \text{ and } v'''_i = \kappa_i(v)
}
{
\sigma | \rho \ \loc.f_i := v \to \sigma' [\loc \mapsto \new{C}{\overline{v'''}}] | \textbf{Unit}
}
\qquad
\ntyperule{\Async}{
\{x\} = \{ x \in e\}, \text{ and } \overline{x'} \text{ are fresh}
}
{
(\sigma, \overline{x \mapsto v} | \textbf{async} \ e) \to
(\overline{\updateStore{\sigma}{\textbf{moved}}{v}},
\overline{x' \mapsto v},
a \mapsto \varnothing\ e[\overline{x := x'}]| a)
}
$$
$$
\nreduction{\Recv}{
(\sigma, a \mapsto v,\overline{v'}\ E[\textbf{rec}]) \to 
(\sigma, a\mapsto \overline{v}\ E[v])
}
\qquad
\nreduction{\Send}{
(\sigma, a\mapsto \overline{v}\ e | \textbf{send}\ a \leftarrow v) \to (\textbf{moved}_v(\sigma), a\mapsto \overline{v},v\ e | \textbf{Unit})
}
\qquad
\nreduction{\Variable}{\sigma | x \to \sigma | \sigma(x)}
$$
$$
\ntyperule{\MethodC}{
\sigma(l) = \new{C}{\ldots}\\
\Delta(C) = \textbf{class}\ C(\ldots) \{ \kappa'\ \textbf{method}\ m(\overline{x: \kappa}) \to \kappa'' \{ e \}\\
\text{for fresh } \overline{x'}, \overline{x''}, \text{ and } \overline{x'''}
}
{
(\sigma | (\rho \ \loc).m(\overline{v})) \to 
		(\kappa'_{\rho l}(\overline{\kappa_{v}}(\sigma)),
		                                      \overline{x' \mapsto \kappa(v)},\
                                                       x'''\mapsto \kappa'(\rho\ \loc)
                                                       | \kappa''\ x'' = e[\overline{x:=x'}][\texttt{this}:= x''']; x'')
}
$$
\caption{\label{fig: runtime} Runtime semantics}
\end{figure*}

%Thus, a reduction step in the calculus is written as follows,
%
%\[
%\sigma, t \mapsto \overline{v}\ E[e]| t_i \mapsto \overline{v} e \to \sigma', t \mapsto \overline{v}\ E[e']|t_i \mapsto \overline{v} e
%\]

%The rule above denotes the execution of expression $e \to e'$ in actor $t$, leading to the modification
%of the store from $\sigma \to \sigma'$. 

In the runtime semantics~(Fig.~\ref{fig: runtime}), the creation of a new instance (\NewClass) updates the store by
casting the argument values to the class to the corresponding
field capabilities ($\overline{\updateStore{\sigma}{\kappa}{v}}$) 
obtained from the class table $\Delta$ 
(throwing a runtime error otherwise). This rule also introduces a new location \loc\
to an object reference ($\loc \mapsto \new{C}{\overline{\kappa\ f}}$),
returning a movable reference location, expressed by the runtime as $\movable\ \loc$.
A field access (\FieldA) simply gets the value from the expected location.
%The semantics encode errors (\Err), \ie{} the runtime does not throw errors immediately
%but waits until there is a read of a location that has been uninitialised.
This allows one actor to move an object to a second actor and still allow the
first actor to keep a reference to it as long as it does not make use of 
it.\footnote{This was left to allow future work, such as a reference capability to a \textbf{moved} object that
contains immutable data. Sharing a reference to immutable data is always thread-safe.}
A field assignment updates the store where the updated field $f_i$ will contain 
the new value $v$, where $v$ needs to conform to its corresponding field
capability, $\updateStore{\sigma}{\kappa_i}{v}$. A method call (\MethodC) updates the store
so that method arguments are consistent with their expected capability
%followed by bindings between the actual arguments and their values, the casting
%of $\self{}$ as $\kappa$ (\ie{} $\updateStore{\sigma}{\kappa}{\rho\ \loc}$),
and explicit \textit{alpha}-conversion, returning the end result of the method.

The asynchronous operations are the creation of a new actor, and
the sending and receiving of data. Spawning a new actor with data $e$ (\Async) updates the store by
uninitialising references that get to the reachable object graph of the data sent,
creating a new actor that will execute $e$ ($a \mapsto \varnothing\ e[\overline{x := x'}]$).
An actor may block to process a new message (\Recv) until it receives
a new message. We assume that actors go back
to an event loop that calls $\rec{}$ upon finishing from processing a message.
Sending data to an actor (\Send) is similar to spawning an actor, in that outside
references to the reachable object graph of the data are uninitialised, followed
by placing the message in the message queue of the actor.

\section{Goals and Challenges}

One of our goals is to perform concurrent
programming in this gradually-based reference capability language. Ideally we
only allow capability annotations on references.
%(and types as soon as
%we target static programming languages). % and avoid extra syntax.
We are planning on adding other reference capabilities
%, such as linear and immutable capabilities, 
and creating its type system.

%The gradual guarantee~\cite{RefinedGradualTyping} requires that in a gradually typed language, one can
%remove any type annotation without changing the behaviour of the program.
In a gradually typed language, one can
remove any type annotation without changing the behaviour of the program~\cite{RefinedGradualTyping}.
We originally tried to add a linear capability with destructive read semantics,
which statically guarantees alias freedom,
but this design violates the gradual guarantee.
The following example
initialises a linear variable $x$, then $y$ aliases $x$, to finally perform a method call.

\begin{lstlisting}[style=grace,xleftmargin=8ex]
lin x := C(...)
y := x
foo(x, y)

method foo(lin y, z)
\end{lstlisting}

We thought that implicit borrowing would be the solution and, upon finding a
linear formal parameter, we performed a destructive read nullifying all aliases,
in particular \ec{z = null}.
Upon removal of the \lin{} annotation from method \ec{foo}, 
\ec{z} is not \ec{null} anymore and may affect the behaviour of the program,
breaking the gradual guarantee. 
%There are other cases similar to this one, and care
%must be taken to not break the gradual guarantee.

\section{Related work}

Gradual typing~\cite{OriginGradualTyping1,OriginGradualTyping2,GradualTyping,RefinedGradualTyping}
allows developers to evolve their programs from the dynamic to the typed discipline,
gradually. Once a typing annotation is introduced, removing it should not change
the behaviour of the program. Our work builds on top of the gradual typing literature,
extending the gradual approach to reference capabilities.

Gradual ownership types~\cite{GradualOwnership} abstract
over ownership annotations to allow gradually change the 
object's ownership structure. In contrast, our work differs
in that we do not statically encode the ownership structure,
but adopt a gradual typing approach to allow developers
to gradually add reference capability annotations.
%In this work,
%\lent{} would be \textit{akin} to external uniqueness~\cite{ExternalUniqueness}.

Fennel \textit{et al}~\cite{LinearLambdaCalculus} used a linear calculus and added gradual typing.
In this approach, the linear annotations are always known, while the type may
be dynamic or statically known. One of the main results is that linearity is
orthogonal to gradual typing. In our approach, we are considering
adding linear capabilities which seem to be a subset of the \moved{}
capabilities (explained in the paper). We also believe that gradual reference
capabilities seem to be orthogonal to gradual typing.

The Haskell programming language is going to support linear annotations,
which is mainly used to perform runtime optimisations such as in-place updates
and to forbid aliasing. Linear Haskell~\cite{LinearHaskell} does not use a gradual approach. Instead,
they feature quantification over multiplicities and parameterised arrows,
which allows the common function $map :: \forall p. (a \to_p b) \to [a] \to_p b$
to be able work as $map :: (a \multimap b) \to [a] \multimap b$ and as
$map :: (a \to b) \to [a] \to b$.

Encore~\cite{SFMEncore,EncoreTS} and Pony~\cite{PonyTS} are capability-based
languages that statically ensure data-race freedom. In these languages, all references
need to be annotated to provide such guarantee. For our work, we took inspiration
from their capability-based model and added a small subset of their capabilities.
We plan on adding more capabilities from these languages, such as deeply immutable
capability (\textit{read} capability in Encore), or linear types
(\lin{} and \textit{iso} from Encore and Pony, respectively).

Boyland \textit{et al} describe a system of capabilities for sharing~\cite{CapabilitiesForSharing}.
Based on this abstract model, we propose a policy that tracks ownership for every local object.
When we send an object, for the object and its reachable object graph, if the object has an owner, its
owner asserts $RWI\bar{R}\bar{W}\bar{I}$. If the object does not have an owner,
then it asserts $ORWI\bar{R}\bar{W}\bar{I}$.

%Encoding Ownership Types in Java~\cite{OwnershipJava}.

%Linear types~\cite{PracticalLinearTypes}, Encore~\cite{SFMEncore,EncoreTS}, Pony~\cite{PonyTS},
%Linear Haskell~\cite{LinearHaskell} and its origins~\cite{qualified-types} and linear
%functional programming~\cite{LinearFunProg,LightweightTypeClasses,PracticalAffine,LinearF}

%Blame tracking~\cite{OriginGradualTyping2,BlameCalculus1,BlameCalculus2}.
%Assume in an untype setting, calling a typed function where we pass an untyped callback
%function. Gradual typing draws on research for contracts and delays the checking until the function is called.
%Thus, a cast error occurs inside the typed function when the untyped parameter is applied.
%With blame tracking, we can report on the call the initiated the call to the typed function
%as the problematic one, instead of pointing to the statically typed one.

%Permission type system using gradual typing~\cite{PermissionGradualTyping}.
%
%Possible related word to dynamic capabilities (dynamic data-race detection)~\cite{PermissionRegions}.
%
%Ownership types~\cite{OwnershipTypes} and data-race freedom~\cite{OwnershipTypesRaceFreedom}.

%Heap snapshots to analyse uniqueness, ownership and confinment~\cite{OwnershipHeap,Spencer}.

\section{Conclusion}

We have presented ongoing work on gradual capabilities and 
presented the runtime semantics of an untyped language that prevents
data-races. We are working on the type system and adding more capabilities.
Fennel \textit{et al} pointed out that linearity is orthogonal to gradual typing,
and we believe that gradual capabilities are also orthogonal to gradual typing,
since we are adding them to an untyped language and the addition of types
is orthogonal to reference capabilities.

%% Bibliography
\bibliography{biblio}

%% Appendix
%\appendix
%\section{Appendix}
%
%Text of appendix \ldots

\end{document}